\documentclass{piparticle-final}
\usepackage{graphicx}
\usepackage{amsmath}

\usepackage{cite} 
\usepackage{epstopdf}

\begin{document}

\volume{6}               
\articlenumber{060003}   
\journalyear{2014}       
\editor{G. Martinez Mekler}   
\reviewers{F. Bagnoli, Dipartimento di Fisica ed Astronomia,\\  \mbox{\,}\hspace{35.5mm} Universita degli Studi di Firenze, Italy}
\received{{11} March 2014}     
\accepted{1 August 2014}   
\runningauthor{A. Chacoma \itshape{et al.}}  
\doi{060003}         

\title{Critical phenomena in the spreading of opinion consensus and disagreement}

\author{A. Chacoma,\cite{inst1} \vspace{1em}
D. H. Zanette\cite{inst1,inst2}\thanks{E-mail:
zanette@cab.cnea.gov.ar}}

\pipabstract{ We consider a class of models of opinion formation
where the dissemination of individual opinions occurs through the
spreading of local consensus and disagreement.  We study the
emergence of full collective consensus or maximal disagreement  in
one- and two-dimensional arrays. In both cases, the probability of
reaching full consensus exhibits well-defined scaling properties as
a function of the system size. Two-dimensional systems, in
particular, possess nontrivial exponents and critical points. The
dynamical rules of our models, which emphasize the interaction
between small groups of agents, should be considered as
complementary to the imitation mechanisms of traditional opinion
dynamics. }

\maketitle

\blfootnote{
\begin{theaffiliation}{99}
\institution{inst1} Instituto Balseiro and Centro At\'omico
Bariloche, 8400 San Carlos de Bariloche, R\'{\i}o Negro, Argentina.
\institution{inst2} Consejo Nacional de Investigaciones
Cient\'{\i}ficas y T\'ecnicas, Argentina.
\end{theaffiliation}
}

\section{Introduction}
\label{sect1}

The remarkable regularities observed in many human social phenomena
---which, in spite of the disparate behavior of individual human
beings, emerge as a consequence of their interactions--- have since
long attracted the attention of physicists and applied
mathematicians. Collective manifestations of human behavior have
been mathematically modeled in a variety of socioeconomic processes,
such as opinion formation, decision making, resource allocation,
cultural and linguistic evolution, among many others, often using
the tools provided by statistical physics  \cite{rev}. The stylized
nature of these models emphasizes the identification of the generic
mechanisms at work in human interactions, as well as the detection
of broadly significant features in their macroscopic outcomes. They
provide the key to a deep insight into the common elements that
underlie those processes.

Models of opinion formation constitute a central paradigm in the
mathematical description of social processes from the viewpoint of
statistical physics. Starting in the seventies and eighties
\cite{set1,set2,och1,och2}, much work ---which we cannot aim at
inventorying here, but which has been comprehensibly reviewed in
recent literature \cite{rev}--- has exploited the formal resemblance
between opinion spreading and spin dynamics in order to apply
well-developed statistical techniques to  the analysis of such
models.

The key mechanism driving most agent-based models of opinion
formation is imitation. For instance, in the voter model ---to which
we refer several times in the present paper--- the basic interaction
event consists in an agent copying the opinion of another agent chosen at random from a specified neighborhood. At any given time,
the opinion of each agent adopts one of two values, typically
denoted as $\pm 1$. The voter model can be exactly solved for
populations of agents distributed over regular (hyper)cubic arrays
in any dimension \cite{redner}. For infinitely large populations, it
is characterized by the conservation of the average opinion. In one
dimension, a finite population always reaches an absorbing state of
full collective consensus, all agents sharing the same
opinion. The probability of final consensus on either opinion
coincides with the initial fraction of agents with that opinion, and
the time needed to reach the absorbing state is of the order of  the
population size squared \cite{rev}.

In this paper, we present an introductory analysis of a class of
models where opinion dynamics is driven by the spreading of
consensus and disagreement, rather than by the dissemination of
individual opinions. The basic concept behind these models is that
agreement of individual opinions in a localized portion of the
population may promote the emergence of consensus in the
neighborhood while, in contrast, local disagreement may inhibit the
growth of, or even decrease, the degree of consensus in the
surrounding region. In real social systems, the mechanism of
consensus and disagreement spreading should be complementary to the
direct transmission of opinions between individual agents. In our
models, however, we disregard the latter to focus on the dynamical
effects of the former.

Since the degree of consensus can only be defined for two or more
agents, the spreading of consensus and disagreement engages groups
of agents rather than individuals. Such groups are, thus, the
elementary entities involved in the social interactions
\cite{g1,g11,g12,g2,g3}. We stress that several other social
phenomena --- related, notably, to decision making \cite{g2} and
resource allocation \cite{DH}--- are also based on group interactions
that cannot be reduced to two-agent events. In the class of models
analyzed here, each interaction event is conceived to occur between
two groups: an {\it active} group $G$ and a {\it reference}
group $G'$. As a result of the interaction, the agents in $G$ change
their individual opinions in such a way that the level of consensus
in $G$ approaches that of $G'$.  This generic mechanism extends
dynamical rules where the opinion of each single agent changes in
response to the collective state of a reference group
\cite{rev,g11,sz1,stauf}. The size and internal structure of the
interacting groups, as well as the precise way in which opinions are
modified in the active group with respect to the reference group,
defines each model in this class. For the sake of concreteness, we
limit the analysis to systems where, as in the voter model,
individual opinions can adopt two values ($\pm 1$). In the next
section, we analyze the case where both the active group and the
reference group are formed by two agents, and the population is
structured as a  one-dimensional array. In this case, the system
admits stationary absorbing states of full consensus and maximal
disagreement, with simple scaling laws with the population size. In
Section \ref{sect3}, we study a two-dimensional version of the same
kind of model with larger groups, where nontrivial critical
phenomena ---not present in the one-dimensional case--- emerge.
Results and perspectives are summarized in the final section.

\section{Two-agent groups on one-dimensional arrays}
\label{sect2}

\begin{figure}[h]
\begin{center}
\includegraphics[width=\columnwidth]{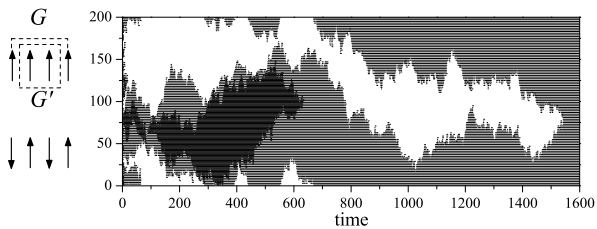}
\end{center}
\caption{Left: The two possible outcomes of the interaction, up to
opinion inversions, for four consecutive agents along the
one-dimensional array. The active and the reference groups, $G$ and
$G'$, are respectively formed by the outermost and innermost agents.
Right: Time evolution of a 200-agent  array with $n_+(0)=0.5$ and
$p_D=p_C=1$. Black and white dots correspond, respectively, to
opinions $+1$ and $-1$. At time $t=1534$, an absorbing state of
maximal disagreement is reached.} \label{fig01}
\end{figure}

We begin by considering the simple situation where each of the two
groups involved in each interaction event is formed by just two
agents. The situation within each group, thus, is one of either full
consensus (when the two agents bear the same opinion, either $+1$ or
$-1$) or full disagreement (when their opinions are different). We
take a population where agents are distributed on a one-dimensional
array, and consecutively labeled from $1$ to $N$. Periodic boundary
conditions are applied at the ends. At each time step,  we choose
four contiguous agents, say, $i-1$ to $i+2$. The central pair $i$,
$i+1$ acts as the reference group $G'$. If they are in disagreement,
the agents $i-1$ and $i+2$ respectively adopt the opinions opposite
to those of $i$ and $i+1$ with probability $p_D$, while with  the
complementary probability $1-p_D$ nothing happens. If, on the other
hand,  $i$ and $i+1$ agree with each other, $i-1$ and $i+2$ copy the
common opinion in $G'$ with probability $p_C$, while with
probability $1-p_C$  nothing happens. In this way, both consensus
and disagreement spread from $G'$ outwards, to the left and right.
The probabilities $p_C$ and $p_D$ control the relative frequency
with which consensus and disagreement are effectively transmitted.
The left panel of Fig.~\ref{fig01} illustrates the states of the
four consecutive agents in the two possible outcomes of the
interaction (up to opinion inversions).

It is not difficult to realize that, for $p_D=p_C=1$, our
one-dimensional array is equivalent to two intercalated
subpopulations ---respectively occupying even and odd sites--- each of
them evolving according to the voter model. The dynamical rules are
reduced in this case to binary interactions between agents. In fact,
whatever the opinions in group $G'$ at each interaction event, agent
$i-1$ and $i+2$ respectively copy  the opinions of $i+1$ and  $i$.
Now, since the voter model always leads a finite population to an
absorbing state of full consensus, the final state of our system can
be one of full consensus on either opinion, or a state of maximal
disagreement where opposite opinions alternate over the sites of the
one-dimensional array. In the latter, the two neighbors of each
agent with opinion $+1$ have opinion $-1$ and vice versa. The right
panel of Fig.~\ref{fig01} shows the evolution of a 200-agent  array
for $n_+(0)=0.5$ and $p_D=p_C=1$, black and white dots respectively
corresponding to opinions $+1$ and $-1$. At any given time, the
population is divided into well-defined domains either of consensus
in one of the opinions or disagreement. Note that the domain
boundaries show the typical diffusive motion found in stochastic
coarsening processes \cite{coars,rev}.

\begin{figure}
\begin{center}
\includegraphics[width=.9\columnwidth]{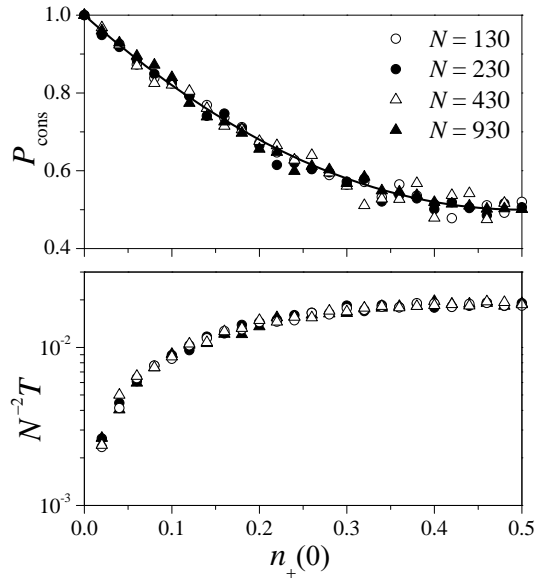}
\end{center}
\caption{Numerical results for consensus and disagreement spreading
on a one-dimensional array with $p_D=p_C=1$, obtained from $10^3$
realizations for each parameter set (see text for details). Upper
panel: Probability of reaching full consensus, $P_{\rm cons}$, as a
function of the initial fraction of agents with opinion $+1$,
$n_+(0)$, for four values of the population size $N$. Lower panel:
Total time $T$ needed to reach the final absorbing state, normalized
by the squared population size $N^2$. Since both $P_{\rm cons}$ and
$N^{-2}T$ are symmetric with respect to $n_+(0)=1/2$, only the lower
half of the horizontal axis is shown.} \label{fig02}
\end{figure}

Taking into account that, in the voter model, the probability of
ending with full consensus on opinion $+1$ is given by the initial
fraction of agents with that opinion, $n_+(0)$, and assuming that
the initial distribution of opinions is homogeneous over the array,
the probability that our system ends in a state of full consensus on
either opinion is $P_{\rm cons}=n_+^2 (0) + n_-^2 (0) = 1-2
n_+(0)+2n_+^2(0)$. Note that this coincides with the probability
that, in the initial state, any two contiguous agents are in
consensus. Moreover, we know that the time needed to reach an
absorbing state in the one-dimensional voter model is proportional
to $N^2$, a result that should also hold in our case.

The upper panel of Fig.~\ref{fig02} shows numerical results for the
probability of final full consensus $P_{\rm cons}$, determined as
the fraction of realizations that ended in full consensus out of
$10^3$ runs, as a function of $n_+(0)$ and for several population
sizes $N$. The curve is the analytic prediction given above. The
result is analogous to the probability of final consensus found in
Sznajd-type models \cite{sz1}.  The lower panel shows the total time
$T$ needed to reach the final absorbing state (of either consensus
or disagreement), averaged over $10^3$ realizations and normalized
by $N^2$. As expected, both $P_{\rm cons}$ and $N^{-2}T$ are
independent of the population size.

\begin{figure}
\begin{center}
\includegraphics[width=.9\columnwidth]{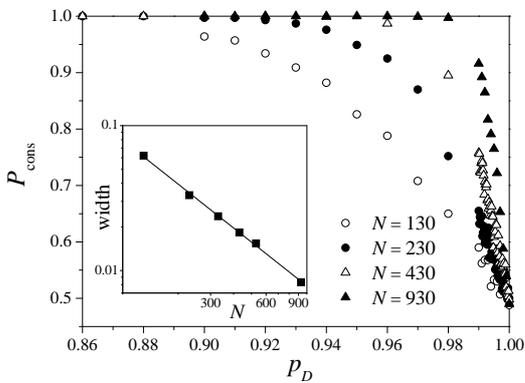}
\end{center}
\caption{Probability of reaching full consensus, $P_{\rm cons}$, as
a function of the probability $p_D$, with $p_C=1$ and for four
values of the system size $N$. Results were obtained averaging over
$10^3$ realizations for each parameter set. Insert: Width of the
variation range of $P_{\rm cons}$ as a function of $N$. The straight
line has slope $-1$. } \label{fig03}
\end{figure}

When $p_D \neq p_C$, the two intercalated subpopulations cannot be
considered independent of each other any more. If $p_D <p_C$, for
instance, an opinion prevailing in one of the subpopulations will
invade the other subpopulation faster than the opposite opinion,
thus favoring the establishment of collective consensus. To analyze
this asymmetric situation, we first fix $p_C =1$ and let $p_D$ vary
in $(0,1)$, so that the spreading of consensus is more probable than
that of disagreement. The main plot in Fig.~\ref{fig03} shows
numerical results for  $P_{\rm cons}$, measured as explained above,
as a function of $p_D$ and for four values of $N$. In all the
realizations, $n_+(0)=0.5$, and the two opinions are homogeneously
distributed over the population. As $p_D$ decreases below $1$, the
probability of reaching full consensus grows rapidly, approaching
$P_{\rm cons}=1$. As $N$ grows, moreover, the change in $P_{\rm
cons}$ is more abrupt. Fitting of a sigmoidal function to the data
of $P_{\rm cons}$ vs.~$p_D$ near $p_D=1$ makes it possible to assign
a width to the range where $P_{\rm cons}$ changes between $1$ and
$0.5$. The insert of Fig.~\ref{fig03} shows this width as a function
of the system size $N$ in a log-log plot. The slope of the linear
fitting is $-1.00 \pm 0.02$. Therefore, the width is inversely
proportional to $N$.

\begin{figure}
\begin{center}
\includegraphics[width=.9\columnwidth]{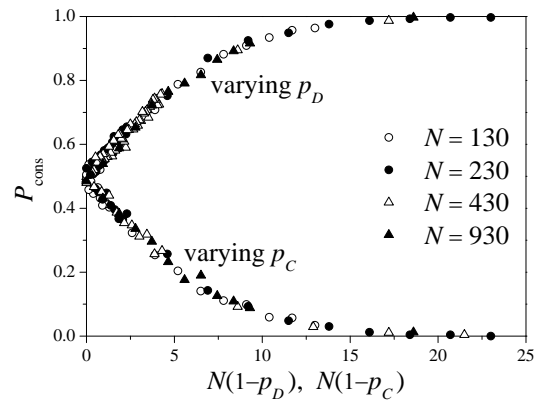}
\end{center}
\caption{Probability of reaching full consensus, $P_{\rm cons}$, as
a function of $N(1-p_D)$ when varying $p_D$ with $p_C=1$, and as a
function of $N(1-p_C)$ when varying $p_C$ with $p_D=1$.}
\label{fig04}
\end{figure}

The facts that $P_{\rm cons} = 0.5$ for $p_D=1$ and for all $N$, and
that the width of the range where $P_{\rm cons}$ changes decreases
as $N^{-1}$, make it possible to conjecture the existence of a
function $\Phi (u)$, with $\Phi (0)=0.5$ and $\Phi (u) \to 1$ for
large $u$, such that $P_{\rm cons} = \Phi[N(1-p_D)]$. To test this
hypothesis, we have plotted our numerical data
for $P_{\rm cons}$ against $N(1-p_D)$ in Fig.~\ref{fig04}. The results are those in the
upper half of the plot (``varying $p_D$''). The collapse of the data
for different $N$ on the same curve confirms the conjecture.

Analogous results were obtained when fixing $p_D=1$ and $p_C$ was
varied. Now, $P_{\rm cons}$ drops to $0$ in a narrow interval for
$p_C$ just below $1$, indicating the prevalence of disagreement.
Again, the width of the interval is proportional to $N^{-1}$. The
results in the lower half of Fig.~\ref{fig04} (``varying $p_C$'')
illustrate the collapse of the corresponding values of $P_{\rm
cons}$ when plotted against $N(1-p_C)$.

\begin{figure}
\begin{center}
\includegraphics[width=.9\columnwidth]{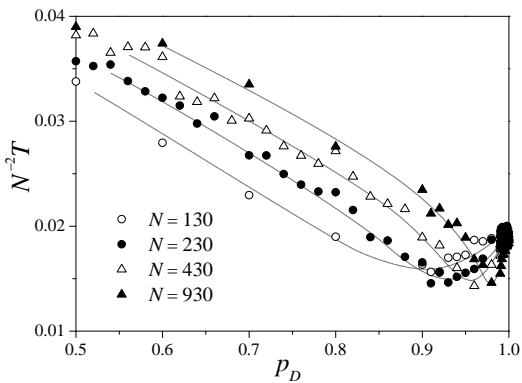}
\end{center}
\caption{Total time $T$ needed to reach the final absorbing state,
normalized by the squared population size $N^2$, as a function of
the probability $p_D$ ($p_C=1$). B\'ezier curves have been plotted
as a guide to the eye.} \label{fig05}
\end{figure}

In our numerical realizations with $p_D \neq p_C$, we have also
recorded the average time $T$ needed to reach the final absorbing
state. Figure \ref{fig05} shows results for $N^{-2} T$ in the case
where $p_C=1$ and $p_D$ changes (cf.~lower panel of
Fig.~\ref{fig02}). In contrast with the case with $p_D=p_C=1$,
rescaling of the time $T$ with $N^2$ leaves a remnant discrepancy
between results for different population sizes $N$. Specifically,
for $p_D<1$, $T$ grows faster than $N^2$. Moreover, $T$ is
nonmonotonic as a function of $p_D$, exhibiting a minimum which
shifts towards $p_D=1$ as $N$ grows. The same dependence with $N$
and $p_C$ is observed when we fix $p_D=1$ and let $p_C$ vary.

Summarizing our results for a one-dimensional population with
two-agent groups, we can say that the possibility that both
consensus and disagreement spread over the system makes it possible
to find absorbing collective states of either full consensus, with
all the agents having the same opinion, or maximal disagreement,
where opposite opinions alternate between consecutive neighbor
agents. For large populations, the relative prevalence of collective
consensus and disagreement is controlled by how the probabilities
$p_D$ and $p_C$ compare with each other. Our results suggest that, in
the limit $N\to \infty$, the condition $p_C > p_D$  univocally leads
to full consensus and vice versa. For smaller sizes, however, the
system can approach full consensus even when $p_D > p_C$, and vice
versa ---presumably due to finite-size fluctuations.

\section{Larger groups on two-dimensional arrays}
\label{sect3}

\begin{figure}
\begin{center}
\includegraphics[width=\columnwidth]{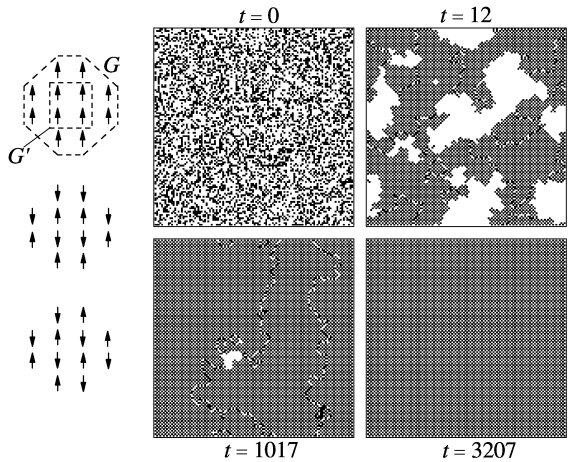}
\end{center}
\caption{Left: The three possible outcomes of the interaction, up to
$\pm 90^\circ$ rotations and opinion inversions, on the
two-dimensional lattice. The active and the reference groups, $G$
and $G'$, are respectively formed by the outermost and innermost
agents. Right: Four snapshots of a  population with $L=120$, for
$n_+(0)=0.35$ and $p_D=p_C=1$, including the initial condition and
two intermediate states. At time $t=3207$, an absorbing state of
maximal disagreement has been reached. Black and white dots
correspond, respectively, to opinions $+1$ and $-1$.} \label{fig06}
\end{figure}

A two-dimensional version of the above model, where agents occupy
the $N=L \times L$ sites of a regular square lattice with periodic
boundary conditions, can be defined as follows. The reference group
$G'$ at each interaction event is a randomly chosen $2 \times
2$-agent block. The corresponding active group $G$ is formed by the
eight nearest neighbors to the agents in $G'$ which are not in turn
members of the reference group. The active group, thus, surrounds
$G'$. Of the sixteen possible opinion configurations of the
reference group, two correspond to full consensus ---with the four
agents sharing the same opinion--- and six correspond to maximal
disagreement ---with two agents in each opinion. The remaining eight
configurations correspond to partial consensus, with only one agent
disagreeing with the other three. The dynamical rules are the
following: (1) if $G'$ is in full consensus, all the agents in $G$
copy the common opinion in $G'$; (2) if $G'$ is in maximal
disagreement, each agent in $G$ adopts the opinion opposite to that
of the nearest neighbor in $G'$; (3) otherwise, nothing happens.
Hence, both consensus and disagreement spread outwards from the
reference group. Probabilities $p_D$ and $p_C$ for the spreading of
disagreement and consensus are introduced exactly as above. The left
part of Fig.~\ref{fig06} shows, up to rotations and opinion
inversions, the three possible outcomes of a single interaction
event.

The states of full collective consensus ---with all the agents in the
population having the same opinion--- and of maximal collective
disagreement ---with the two opinions alternating site by site along
each direction over the lattice--- are absorbing states, in
correspondence with the one-dimensional case. However, for
$p_D=p_C=1$, the system cannot be reduced anymore to a collection of
sublattices governed by the voter model. The definition of $G$ and
$G'$ establish now correlations between the  opinion changes in the
active group at each interaction event. Moreover, some opinion
configurations in the reference group induce evolution in the active
group, while others do not. Figure \ref{fig06} shows, to its right,
four snapshots of a $120 \times 120$-agent population, along a
realization starting with $n_+(0)=0.35$ and $p_D=p_C=1$. Note the
formation of consensus clusters at rather early stages, and the
final prevalence of disagreement. The line boundaries between
disagreement regions are also worth noticing.

\begin{figure}
\begin{center}
\includegraphics[width=.9\columnwidth]{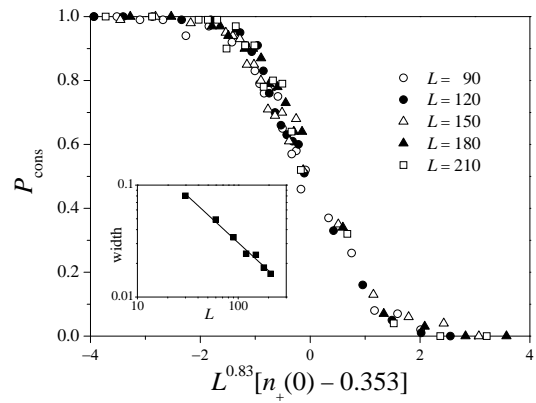}
\end{center}
\caption{Numerical results for the probability of reaching full
consensus, $P_{\rm cons}$, on a two-dimensional lattice with
$p_D=p_C=1$, obtained from $100$ realizations for each parameter
set. Collapse for several system sizes $L$ is obtained plotting
$P_{\rm cons}$ against $L^{0.83} [n_+(0)-0.353]$.  Insert: Scaling
of the width of the transition zone of $P_{\rm cons}$, determined
from fitting a sigmoidal function, as a function of the size $L$.
The straight line has slope $-0.83$.} \label{fig07}
\end{figure}

Following the same lines as for the one-dimensional array, we study
first the probability $P_{\rm cons}$ of reaching full collective
consensus as a function of the initial fraction of agents with
opinion $+1$, $n_+(0)$, in the case $p_D=p_C=1$. Opinions are
homogeneously distributed all over the population. For very small
$n_+(0)$, as expected, we find $P_{\rm cons}\approx 1$. However, in
sharp contrast with the one-dimensional case (see Fig.~\ref{fig03}),
$P_{\rm cons}$ remains close to its maximal value until $n_+ (0)
\approx 0.35$, where it drops abruptly to $P_{\rm cons}\approx 0$.
The width of the transition zone decreases as a nontrivial power of
the system size, $\sim L^{0.83\pm 0.04}$, as illustrated in the
insert of Fig.~\ref{fig07}. Our best estimate for the critical value
of $n_+ (0)$ at which $P_{\rm cons}$ drops is $n_+^{\rm crit} =
0.353 \pm 0.001$. The main plot in the figure shows the collapse of
numerical measurements of $P_{\rm cons}$ as a function of $n_+(0)$
for different sizes $L$, averaged over $100$ realizations, when
plotted against the rescaled shifted variable $L^{0.83} [n_+(0) -
0.353]$.

These results suggest that, for very large populations, the
probability of reaching full consensus jumps discontinuously from
$P_{\rm cons} =1$ to $0$ at $n_+(0) = n_+^{\rm crit}$. Compare this with
the smooth, size-independent behavior of the one-dimensional case.
Note also that $n_+^{\rm crit}$ is close to, but does not coincide
with, $n_+(0) =1/3$. At this latter value, in the initial condition
with homogeneously distributed opinions, the probability of finding
a $2 \times 2$-agent block in full consensus becomes lower than that
of maximal disagreement as $n_+(0)$ grows.

\begin{figure}
\begin{center}
\includegraphics[width=.9\columnwidth]{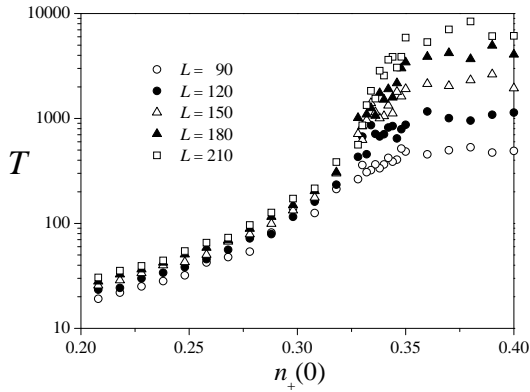}
\end{center}
\caption{Total time $T$ needed to reach the final absorbing state in
a two-dimensional lattice, as a function of $n_+(0)$, for different
sizes $L$.} \label{fig08}
\end{figure}

In the above simulations, we have also measured the average total
time $T$ needed to reach the final absorbing state. Results are
shown in Fig.~\ref{fig08}. Again in contrast with the
one-dimensional case, $T$ exhibits a remarkable change in its
scaling with the system size as $n_+(0)$ overcomes the critical
value $n_+^{\rm crit}$.

\begin{figure}
\begin{center}
\includegraphics[width=.9\columnwidth]{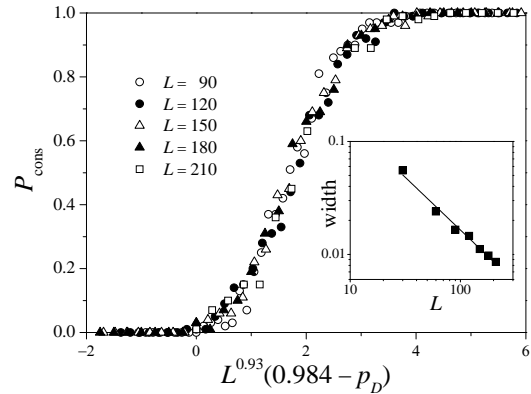}
\end{center}
\caption{Collapse of numerical results for the probability of
reaching full consensus, $P_{\rm cons}$, on a two-dimensional
lattice with $p_C=1$ and $n_+(0)=0.5$, for several system sizes $L$
when plotted against $L^{0.93} (0.984-p_D)$. Insert: Scaling of the
width of the transition zone of $P_{\rm cons}$ as a function of the
size $L$. The straight line has slope $-0.93$.} \label{fig09}
\end{figure}

Going now to the dependence of $P_{\rm cons}$ on the probability of
disagreement spreading $p_D$ ---with $p_C=1$ and $n_+(0)=0.5$--- it qualitatively mirrors that of the
one-dimensional case, shown in Fig.~\ref{fig03}. Namely, as $p_D$
decreases from $1$, $P_{\rm cons}$ grows from $0$ to $1$ in an
interval whose width decreases with the population size. In the
two-dimensional system, however, the transition takes place at a
critical probability $p_D^{\rm crit}$ that can be clearly discerned
from $p_D=1$. Our estimate is $p_D^{\rm crit} = 0.984 \pm 0.002$.
Moreover, the scaling of the transition width with the population
size exhibits a nontrivial exponent, decreasing as $L^{- 0.93 \pm
0.05}$. Collapse of the rescaled numerical results for various
sizes, obtained from averages of $100$ realizations, are shown in
Fig.~\ref{fig09}, where we plot  $P_{\rm cons}$  as a function of
$L^{0.93} (0.984-p_D)$ (cf.~Fig.~\ref{fig04}). The insert displays
the power-law dependence of the width on the size $L$. Analogous
results are obtained if the probability of consensus spreading $p_C$
is varied, with $p_D=1$.

\begin{figure}
\begin{center}
\includegraphics[width=.9\columnwidth]{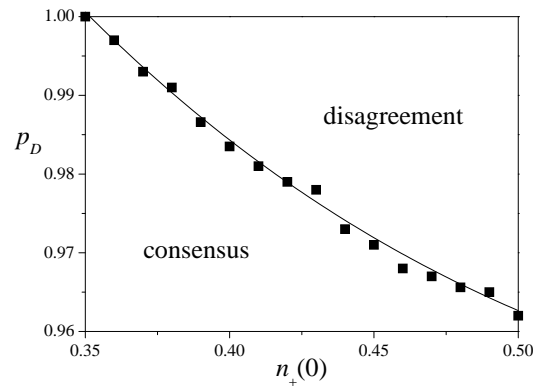}
\end{center}
\caption{Zones of relative prevalence of full consensus and maximal
disagreement in a two-dimensional lattice with $L=120$, plotted on
the parameter plane $(n_+(0),p_D)$. Symbols stand for numerical
results, and the curve serves as a guide to the eye.} \label{fig10}
\end{figure}

Finally, we have found that the transition in $P_{\rm cons}$ as a
function of the disagreement probability $p_D$ shows a dependence on
the initial fraction of agents with opinion $+1$. To characterize
this effect in a way that highlights the relative prevalence of
disagreement and consensus, we have measured the value of $p_D$ at
which the probability of getting full collective consensus reaches
$P_{\rm cons}=0.5$, as a function of $n_+(0)$. The parameter plane
$(n_+(0),p_D)$, thus, becomes divided into regions where a final state
of full consensus is more probable than that of maximal
disagreement, and vice versa. Results for a $120 \times 120$-agent
population are presented in Fig.~\ref{fig10}.

In summary, while spreading of consensus and disagreement on a
two-dimen\-sional lattice bears  superficial qualitative similarity
with the one-dimensional case, the probability that the population
reaches full collective consensus in two dimensions exhibits a quite
different dependence on the system size, on the initial conditions,
and on the spreading probabilities. In particular, our results
reveal the existence of critical phenomena involving scaling laws
with nontrivial exponents.

\section{Conclusion}

In this paper, we have considered the emergence of collective
opinion in a population of interacting agents where, instead of
imitation between individual agents, opinions are transmitted
through the spreading of local consensus and disagreement  toward
their neighborhoods. The basic interacting units in this mechanism
are not individual agents but rather small groups of agents which
mutually compare their internal degrees of consensus and modify
their opinions accordingly.  In this sense, it extends the basic
mechanism underlying such models as the majority-rule and
Sznajd-like dynamics \cite{rev,g11,sz1}, where the opinion of each
individual agent changes in response to the collective state of a
reference group. It is expected that in real social systems the
dissemination of individual opinions through agent-to-agent
imitation on one side, and the spreading of consensus and
disagreement by group interaction on the other, are complementary
mechanisms simultaneously shaping the overall opinion distribution.
Here, in order to gain insight on the specific effects of the second
class, we have focused on models solely driven by the spreading of
consensus and disagreement. The combined effects of the two
mechanisms is a problem open to future work.

Our numerical simulations concentrated on two-opinion models
evolving on one- and two-dimensional arrays \cite{stauf}. In both
cases, absorbing states with all the population bearing the same
opinion (full consensus) and with half of the population in each
opinion (maximal disagreement) are possible final states for the
system. Maximal disagreement states are characterized by alternating
opinions between neighbor sites along the arrays.

A relevant quantity to characterize the behavior is the probability
of reaching full consensus, as a function of the initial condition
---i.e., the initial fraction of the population with each opinion---
and of the relative probabilities of consensus and disagreement
spreading. The total time needed to reach the final absorbing state,
averaged over realizations, has also been measured as a
characterization of the dynamics. We have found that, in several
cases, these quantities display critical phenomena when the control
parameters are changed, with power-law scaling laws as functions of
the system size, pointing to the presence of discontinuities in the
limit of infinitely large populations. It is interesting to remark
that the scaling laws are rather simple for one-dimensional arrays,
but involve nontrivial exponents and critical points in the case of
two-dimensional systems.

Within the same one- and two-dimensional models analyzed here, an
aspect that deserves further exploration is the dynamics and mutual
interaction of the opinion domains that develop since the first
stages of evolution (Figs.~\ref{fig01} and \ref{fig06}). However,
the most interesting extension of the present analysis should
progress along the direction of considering more complex social
structures. The interplay between the dynamical rules of consensus
and disagreement spreading and the topology of the interaction
pattern underlying the population might bring about the emergence of
new kinds of collective self-organization phenomena.

\begin{acknowledgements}
We acknowledge enlightening discussions with Eduardo Jagla.
Financial support from ANPCyT (PICT2011-545) and SECTyP UNCuyo
(Project 06/C403), Argentina, is gratefully acknowledged.
\end{acknowledgements}

\end{document}